
%
\input harvmac

\def\ts{\tilde{\sigma}}
\def\ut#1{\rlap{\lower1ex\hbox{$\sim$}}#1{}}
\def\sut#1{\rlap{\lower2ex\hbox{$\,\tilde {}$}}#1{}}
\def\sep{\rlap{\lower2ex\hbox{$\,\tilde {}$}}\epsilon{}}
\def\tB{\tilde{B}}

\def\Ttitle#1#2#3{\nopagenumbers\abstractfont\hsize=\hstitle\rightline{#1}%
\nopagenumbers\abstractfont\hsize=\hstitle\rightline{#2}%
\vskip 1in\centerline{\titlefont #3}\abstractfont\bigskip \pageno=0}

\Ttitle{VPI-IHEP-93-5, gr-qc/9307018}{CGPG-93/9-2}{Superspace
Dynamics}
\bigskip
\centerline{{\titlefont Perturbations around ``Emptiness''}}

\vskip 0.5in
\centerline{Chopin Soo\foot{Present address:Center for Gravitational
Physics and Geometry, Penn. State University, University Park,
PA 16802-6300; e-mail:soo@phys.psu.edu}
\& Lay Nam Chang\foot{Present address:Theoretical Physics, NSF,
Washington, DC; e-mail:lchang@nsf.gov}
}
\centerline{Institute for High Energy Physics}
\centerline{Virginia Polytechnic Institute and State University}
\centerline{Blacksburg, Virginia 24061-0435}
\bigskip\bigskip

\centerline{{\bf Abstract}}
\bigskip
Superspace parametrized
by gauge potentials instead of metric three-geometries is discussed
in the context of the Ashtekar variables. Among other things, an
``internal clock'' for the full theory can be identified. Gauge-fixing
conditions which lead to the natural geometrical separation of physical
from gauge modes are derived with the use of the supermetric in
connection-superspace. A perturbation scheme about an unconventional
background which is inaccessible to conventional variables is presented.
The resultant expansion retains much of the simplicity of Ashtekar's
formulation of General Relativity.
\bigskip
\Date{Revised 9/93}


In the ADM formalism[1], the superhamiltonian constraint can be
written as
\eqn\one{
{\cal H}\,\equiv (16\pi G)G_{ijkl}\pi ^{ij}\pi ^{kl}
+\sigma \,{{\sqrt{g}}\over{16\pi G}}\,^3R =0}
where $\sigma $ takes the value of $+1$ for spacetimes of Euclidean
signature and $-1$ for spacetimes of Lorenztian signature. As noted
by the authors of Ref. 2, the theory has an interesting strong coupling
limit or zero signature limit at which the potential term vanishes
and only the kinetic term which is quadratic in the momenta remains.
$G_{ijkl}$ can be assumed to be the metric of superspace
(the space of 3-geometries described by the equivalence classes of
spatial metrics under 3D-diffeomorphisms) and it has the form[3]
\eqn\two{
G_{ijkl}={{1}\over{2\sqrt{\vert g\vert }}}(g_{ik}g_{jl}+
g_{il}g_{jk}\,-\,g_{ij}g_{kl})}%
with inverse
\eqn\three{
G^{ijkl}=\,{{1}\over{2}}\sqrt{\vert g\vert }(g^{ik}g^{jl}+
g^{il}g^{jk}-2g^{ij}g^{kl})}%
The supermetrics are ultralocal in the spatial metric variables.
Moreover, an intrinsic time parameter which is proportional to
ln$ \vert g\vert$ can be identified since the supermetric
\eqn\four{
\delta S^2(\vec x)\,=\,{{1}\over{2}}\sqrt{\vert g\vert }(g^{ik}
g^{jl}+g^{il}g^{jk}-2g^{ij}g^{kl}
)\delta g_{ij}\delta g_{kl}}
has hyperbolic signature ($-,+,+,+,+,+)[3].$ This suggests that
in quantum gravity, especially in the context of spatially compact
manifolds, a preferred degree of freedom of the theory can be singled
out as the ``internal clock'' relative to which other degrees of
freedom of the theory evolve according to the dynamics governed
by the Wheeler-DeWitt Equation. The adoption of such an approach
could lead to a resolution of the issue of time in quantum gravity.
(For a discussion on the ``issue of time" in quantum gravity in the
context of the connection variables we will be focussing on, see
Chap. 12 of Ref. 7). With expression (4) as the metric of superspace,
in the strong coupling (zero signature) limit, the superhamiltonian
constraint can be interpreted to be the free Klein-Gordon equation[2].

Ashtekar has achieved remarkable simplifications of the constraints
of General Relativity by introducing SO(3) gauge potentials as fundamental
variables[4$-7].$ In terms of the new variables, the constraints
for pure gravity read
\eqna\five
$$\eqalignno{
G^a &\equiv D_i{\widetilde \sigma }^{ia}  =0   &\five a \cr
{}
H_i &\equiv \sep_{ijk}{\widetilde B}^{jb}{\widetilde \sigma }^{ia}=0 %
&\five b \cr
{}
H &\equiv    \sep_{ijk}   \epsilon _{abc}{\widetilde B}^{kc}{\widetilde \sigma
}^{ia}{\widetilde \sigma }^{jb}
 =  0  &\five c\cr
}$$
where the magnetic field
\eqn\six{
 {\widetilde B}^{ia}   \equiv {{1}\over{2}}{\widetilde \epsilon }^{ijk}
\left\{ \partial_jA^a_k-\partial_kA^a_j+{\epsilon^a}_{bc}A^b_j
 A^c_k\right\}
 }
The tildes above and below the variables denote the fact that they
are tensor densities of weight $+1$ and $-1$ respectively.
In the above, lower case Latin indices from $a$ to $c$
 denote internal SO(3) indices while indices from $i$
 onwards are spatial indices. All these indices run from
1 to 3. We can replace $\epsilon _{abc}$ by the tensor density
$\sep_{abc}$ in (5c) so that rather than being
of weight two, the superhamiltonian is of weight one as is the case
for ${\cal H}$ with the ADM variables. This makes the supermetric
in connection superspace, expression (16) below, gauge and
3D-diffeomorphism-invariant without the introduction of metric
or triad variables.

In particular, among the simplifications achieved by Ashtekar, there
is remarkably no potential term in the superhamiltonian constraint
of the {\it full} theory, if we treat ${\widetilde \sigma }^{ia}$
as the momentum variable, and adopt the natural choice of
\eqn\ {
 {\ut{G}}_{iajb}\equiv    \sep_{ijk}
 \sep_{abc}{\widetilde B}^{kc}
}
as the contravariant metric for the space of ``gauge-invariant
3-geometries"
described by the equivalence classes of Ashtekar connections under
gauge transformations and 3D-diffeomorphisms. This definition of
the supermetric does not involve the variables ${\widetilde \sigma }^{ia}$
or $g_{ij}$ . In terms of vielbeins
\eqna\eight
$$\eqalignno{
 {\ut{G}}_{iajb}&\equiv {{1}\over{3}}  ( {e^c}_c)_{ia}({e^d}_d)_{jb}+
( {{\bar e}^c}_d)_{ia}({{\bar e}^d}_c)_{jb}  &\eight a \cr
{}\quad
  &\equiv   ( e^0)_{ia}(e^0)_{jb}-  ( e{'^c}_d)_{ia}(e{'^d}_c)_{jb} %
&\eight b \cr
}$$
where the vielbeins can be taken to be
\eqna\nine
$$\eqalignno{
 \sqrt{{{1}\over{3}}}({e^c}_c)_{ia}&=(e^0)_{ia}=({{2}\over{3}} %
{\tilde B})^{1/2}   {\sut{b}}{}_{ia} &\nine a \cr
{}
 \pm i({{\bar e}^c}_d)_{ia}&=(e{'^c}_d)_{ia}={\widetilde B}^{{{1}
\over{2}}}\left\{\pm {{1}\over{3}}   {\sut{b}}{}_{ia}{
\delta ^c}_d\mp    {\sut{b}}^c{}_i\delta _{ad}\right\} &\nine b \cr
 }$$
Here ${\sut{b}}{}_{ia}$ denotes the inverse of the magnetic
field and
\eqn\ {
{\tilde B} \equiv    {{1}\over{3!}}   \sep_{ijk
}\sep_{abc}{\widetilde B}^{ia}{\widetilde B}^{jb}{\widetilde B}^{kc} %
 }
is the determinant of ${\widetilde B}^{ia}.$ Notice however, that
unlike expression (2), the supermetric (8) is not ultralocal in
$A$. See Ref. 8 for some comments on the trade off between an ultralocal
supermetric with a local potential term in the ADM formalism and
a local supermetric without any potential in the Ashtekar formalism.
The covariant supermetric, the inverse of ${\ut{G}}_{iajb}$
in the Ashtekar formalism, is readily computed to be
\eqn\ {
 {\widetilde G}^{iajb} \equiv    {\widetilde B}^{-1}\left\{
 {{1}\over{2}}   {\widetilde B}^{ia}{\widetilde B}^{jb}   -   {\widetilde
B}^{ib}{\widetilde B}^{ja}
\right\}
 }
with
\eqn\ {
 {\widetilde G}^{iakc}(\vec x){\ut{G}}_{kcjb}(\vec x)
 ={\ut{G}}_{jbkc}(\vec x){\widetilde G}^{kcia}(\vec x)
 =   {\delta ^i}_j{\delta ^a}_b
 }
$\vec x$ denotes the coordinate on an initial-value hypersurface.
Since
\eqn\ {
\det( {\ut{G}}_{iajb})=\left\{ \det({\widetilde G}^{iajb})
\right\} ^{-1}=-2{\widetilde B}^3
}
the inverse (covariant) supermetric exists if and only if the magnetic
field is non-degenerate. In this report, unless stated otherwise,
we shall deal only with Ashtekar variables that are real and side-step
the reality conditions[4$-7]$ that have to be imposed on the Ashtekar
variables. For space-times with Euclidean signature, it is consistent
to assume that all the variables are real.

In superspace parametrized by gauge potentials, the supermetric
takes the form
\eqna\fifteen
$$\eqalignno{
 &\delta S^2(\vec x)   =   {\widetilde G}^{iajb}(\vec x)\delta A_{ia}(\vec x)
\delta A_{jb}(\vec x)  &\fifteen a \cr
{}
&=   {\widetilde B}^{-1}\left\{{{1}\over{2}}
({\widetilde B}^{ia}\delta A_{ia})({\widetilde B}^{jb}
\delta A_{jb})-({\widetilde B}^{ia}\delta A_{ib})({\widetilde B}^{jb}
\delta A_{ja})\right\} &\fifteen b \cr
{}
&=   {\widetilde B}^{-1}\left\{   {{1}\over{6}}(\delta {X^a}_a)^2
 -\delta {{\bar X}^a}_b\delta {{\bar X}^b}_a\right\}  &\fifteen c \cr
 }$$
where the {\it local} coordinates ${X^a}_b$ are defined
by
\eqn\ {
\delta {X^a}_b   \equiv {\widetilde B}^{ia}\delta A_{ib};\qquad \delta
{{\bar X}^a}_b   \equiv    \delta {X^a}_b-   {{1}\over{3}}   {\delta
^a}_b(\delta {X^c}_c)
 }
Curvature of superspace obstructs the integrability of the local
coordinates. The supermetric
\eqn\ {
 \delta S^2=\int    d^3\vec x   {\widetilde G}^{iajb}(\vec x)\delta
A_{ia}(\vec x)\delta A_{jb}(\vec x)
}
is clearly diffeomorphism and SO(3) gauge-invariant and does not
require the metric variables ${\widetilde \sigma }^{ia}$ or $g_{ij}$
for it to be defined. The supermetric has signature
\eqn\ {
\hbox{\rm sign}( {\widetilde B}^{-1})(+,+,+,+,-,-,-,-,-)
}
since a straightforward decomposition of $\delta $\=X into symmetric
traceless and anti-symmetric parts
\eqn\ {
 \delta {{\bar X}^a}_b   \equiv \delta {S^a}_b+   {{1}\over{2}}{
\epsilon ^a}_{bc}\delta T^c
}
yields
\eqna\twenty
$$\eqalignno{
 &\delta S^2(\vec x)   =   {\widetilde B}^{-1}\left\{   {{1}\over{6}}(
\delta {X^a}_a)^2   -\delta {{\bar X}^a}_b\delta {{\bar X}^b}_a\right\}
&{} \cr
{}
 &=   {\widetilde B}^{-1}\left\{   {{1}\over{6}}(\delta {X^a}_a)^2
 +   {{1}\over{2}}(\delta T^c)(\delta T_c)\right\}  &{} \cr
{}
&\quad -\left\{2(\delta {S^1}_2)^2+2(\delta {S^2}_3)^2
          +2(\delta {S^3}_1)^2+{{3}
\over{2}}(\delta {S^1}_1+\delta {S^2}_2)^2+{{1}\over{2}}(\delta
{S^1}_1-\delta {S^2}_2)^2  \right\} &\twenty{}  \cr
 }$$
It will be shown that when restricted to the physical subspace modulo
the constraint (5b), the hyperbolic supermetric has signature
sign(${\widetilde B}^{-1})(+,-,-,-,-,-)$
and the anti-symmetric matrices ${\epsilon ^a}_{bc}\delta T^c$ drop out.

The factor ${\widetilde B}^{-1}$ would have been a conformal factor
in $\delta S^2(\vec x)$ were it to be positive definite. We make
no such restrictions;  but note that the signature of the supermetric
is determined by the sign of ${\widetilde B}^{-1}$, so that there
is a switching of time-like and space-like coordinates in connection-superspace
when $\tB$ reverses sign. This situation is much akin to what happens
in spacetime when one crosses the horizon of a black hole. This
parallel is more than a mere analogy. In superspace, the crossover
in the sign of $\tB$ occurs at vanishing $\tB$. A computation of the
Ashtekar connection one-forms for the classical Schwarzschild solution
yields[9]
\eqna\twone
$$\eqalignno{
 A_{1} &=   \pm    {m\over r^2}  idt  \pm   \cos \theta d\phi
       &\twone a \cr
{}
 A_{2}   &=   -\sqrt{1-{2m\over r}} \sin \theta d\phi  &\twone b \cr
{}
 A_{3}   &=  \sqrt{1 -{2m\over r}}d \theta   &\twone c \cr
}$$
if the vierbein one-forms ($g_{\mu \nu }dx^\mu dx^\nu =e_{A}$
$e_{A})$ are taken to be
\eqn\ {
 e_{A}   =  \left\{  \pm \sqrt{1-{2m\over r}}i dt ,
{dr\over \sqrt{1 -{2m\over r}}} ,
rd \theta   , r\sin \theta d\phi   \right\}
 }
(As emphasized in Refs. 9, the classical Ashtekar connections are
the anti-self-dual part of the spin connection and therefore depends
on the orientation of the vierbeins). It can be seen that at the
horizon r$=2m,$ the Ashtekar connection becomes abelian and $\tB$ vanishes.
Moreover, it can be checked that $\tB$ remains real and changes sign
when we cross the horizon. It is appropriate here to interject a
word of caution. The expressions (20) and (21) are for the Schwarzschild
solution with Lorenztian signature. To obtain the Euclidean Schwarzschild
solution, we can make the continuation t $\rightarrow -i\tau ,$
but the Euclidean Schwarzschild solution exists only for r$\geq
2m$$[10]$,  which also follows from requiring the Ashtekar
potentials to be real. $\tB$ still vanishes at the horizon of course.
In general, for real connections, the gauge-invariant condition
$\tB(\vec x)=0$ has codimension one and thus naturally defines a
two-dimensional surface on which it holds.  (For complex Ashtekar
connections which corresponds to spacetimes with Lorentzian signature,
reality conditions have to be imposed on the variables but as we
have discussed, for the Schwarzschild solution, $\tB$ of the Ashtekar
connection remains real and switches sign at the horizon).

The covariant supermetric in connection superspace exists if and
only if $\tB$ is non-vanishing. Thus configurations with vanishing
$\tB$ correspond to points in superspace at which the supermetric
(9) is singular. Some examples are manifolds with horizons and manifolds
described by Ashtekar potentials which are abelian anti-instantons[9].
In non-perturbative quantum gravity, such configurations stand out
and can be expected to play crucial roles.

The Gauss' Law and supermomentum constraints are first order in
the momenta and generate SO(3) gauge transformations and gauge-covariant
3D diffeomorphisms respectively. Their constraint algebra closes
without structure functions which depend on dynamical variables.
This suggests that these constraints are to be treated differently
from the superhamiltonian constraint which is quadratic in the momenta.
Physically, it can be interpreted that the theory depends only on
gauge-invariant 3-geometries and as is suggested by the form of
the supermetric, the superhamiltonian constraint can then be used
to determine the dynamical evolution of the theory with respect
to an intrinsic time parameter.

The existence of a supermetric allows a local decomposition of the
cotangent space into gauge directions and their orthogonal complement
with respect to the supermetric. There is thus a natural geometrical
separation of the physical from the gauge degrees of freedom through
the derived gauge-fixing conditions. Some recent works on the relevance
of such gauge-fixing conditions in gravity using conventional variables
can be found in Ref. 11. With ${\widetilde G}^{iajb}$ as the supermetric,
the gauge-fixing conditions are obtained from
\eqn\ {
 \int    d^3\vec x   {\widetilde G}^{iajb}\delta A^\perp _{ia}(\delta
A^g_{jb})=0
 }
where $\delta A^\perp $ and $\delta A^g$ are the physical modes and
gauge directions respectively. Using the explicit form of the contraints,
we obtain
\eqn\ {
 \int    d^3\vec x   {\widetilde G}^{iajb}\delta A^\perp _{ia}(-D_j
\eta )_b=0
}
and
\eqn\ {
 \int    d^3\vec x   {\widetilde G}^{iajb}\delta A^\perp _{ia}(
 \sep_{jkl}{\widetilde B}^{kb}\xi ^l)=\int
   d^3\vec x  2 \xi ^l{\epsilon ^{ac}}_bb_{lc}{\widetilde B}^{ib}
\delta A^\perp _{ia}=0
 }
where $b_{ia}$ $\equiv {{1}\over{2!}}$ ${\widetilde B}^{-1}$ ${\sep}{}_{ijk }
\epsilon _{abc}{\widetilde B}^{jb}{\widetilde B}^{kc}$ .

The six gauge-fixing conditions are derived from the requirement
that Eqns. 23 and 24 hold for arbitrary $\eta _a(\vec x)$ and $\xi
^i(\vec x).$ Thus for non-degenerate magnetic fields, the gauge-fixing
conditions for 3D diffeomorphisms lead to
\eqn\ {
 {\epsilon ^{ac}}_b{\widetilde B}^{ib}\delta A^\perp _{ia}=0
 }
In terms of the local coordinates of Eqn. 15, this is the same as
the requirement that $\delta {X^a}_b$ be symmetric. Thus the unphysical
modes $\delta T^a$ can be set to zero and we are left with a supermetric
of signature sign(${\widetilde B}^{-1})(+,-,-,-,-,-)$ which picks
out ${X^a}_a$ as the preferred intrinsic time-like coordinate in
superspace. Notice that the time-like coordinate has the interesting
property of
\eqn\ {
 \int d^3\vec x   \delta {X^a}_a=\delta {\cal C}
}
where ${\cal C}$ is the Chern-Simons functional. It should be noted
that with the Ashtekar variables, it is the supermomentum rather
than the Gauss' Law constraint which eliminates the subspace of
the supermetric which corresponds to the antisymmetric part of ${\widetilde
B}^{ia}
\delta A_{ib}.$ Alternatively, if we order the supermomentum constraint
in the connection-representation as
\eqn\ {
 \sep_{ijk}{\widetilde B}^{ja}   {{\delta
}\over{\delta A_{ka}}}\Phi [X(A)]   =  0
}
and use
\eqn\ {
 {{\delta }\over{\delta A_{ia}(\vec x)}}   \Phi [X(A)]   =\int d^3\vec y
 {{\delta {X^b}_c(\vec y)}\over{\delta A_{ia}(\vec x)}}   {{\delta
\Phi }\over{\delta {X^b}_c(\vec y)}}   ={\widetilde B}^{ib}(\vec x)
 {{\delta \Phi }\over{\delta {X^b}_a(\vec x)}}
}
we have
\eqn\ {
  \sep_{ijk}{\widetilde B}^{ja}   {\widetilde B}^{kb}
\epsilon _{abc}   {{\delta \Phi }\over{\delta T_c}}   =  0
}
which, for non-degenerate magnetic fields, implies $\delta\Phi /{\delta T_c}$
$=$ 0 .

Three other gauge modes which correspond to the SO(3) gauge-invariance
of the Ashtekar variables can be eliminated using (23) which leads
to the gauge-fixing condition
\eqn\ {
[ D_i({\widetilde G}^{i.jb}\delta A^\perp _{jb})]^a=\partial _i({\widetilde
G}^{iajb}
\delta A^\perp _{jb})   +\epsilon _{abc}A^b_i{\widetilde G}^{icjd}
\delta A^\perp _{jd}=0
}
With the non-trivial supermetric, this is a natural generalization.
For ordinary SO(3) gauge theory in flat space-time, this gauge-fixing
condition reduces to the usual covariant Coulomb gauge condition
\eqn\ {
( D^i\delta A^\perp _i)^a=0
 }
since the supermetric for this particular instance is flat and is
of the form ${\widetilde G}^{iajb}=\delta ^{ij}\delta ^{ab}.$ The
gauge-fixing condition (30) however involves the gauge-invariant
intrinsic time parameter relative to which other degrees of freedom
are to evolve. A more reasonable alternative gauge-fixing procedure
is to use the supermetric for the subspace complement to the intrinsic
time coordinate to eliminate the gauge degrees of freedom of the
subspace. Since we can also write
\eqn\ {
( \delta S)^2   =  \left\{ [( E^0)^{ia}\delta A_{ia}]^2
-{\overline{\hbox{G}}}^{iajb}\delta A_{ia}\delta A_{jb}\right\} %
}
the supermetric for the desired subspace has the form
\eqna\thfour
$$\eqalignno{
 {\overline{G}}^{iajb}   &\equiv   ( {{\bar E}^b}_c)^{ia}
({{\bar E}^c}_b)^{jb} &\thfour a \cr
{}
\quad  &={\widetilde B}^{-1}\left\{   {\widetilde B}^{ib}{\widetilde B}^{ja}
 -{{1}\over{3}}   {\widetilde B}^{ia}{\widetilde B}^{jb}\right\}  %
&\thfour b \cr
 }$$
with vielbeins
\eqna\extra
$$\eqalignno{
( E^0)^{ia} &=(6 \tB )^{-{{1}\over{2}}}{\widetilde B}^{ia}  &\extra a \cr
( {{\bar E}^b}_c)^{ia} &={\widetilde B}^{-{{1}
\over{2}}}\left\{\pm {\widetilde B}^{ia}{\delta ^b}_c\mp {{1}
\over{3}}{\widetilde B}^{ib}{\delta ^a}_c\right\}  &\extra b \cr
 }$$
The resultant gauge-fixing conditions from
\eqn\ {
 \int    d^3\vec x   {\overline{G}}^{iajb}\delta A^\perp _{ia}(\delta
A^g_{jb})   =  0
 }
are as before
\eqn\ {
 {\epsilon ^{ac}}_b{\widetilde B}^{ib}\delta A^\perp _{ia}=0 \qquad i.e.
 \delta {{\bar X}^a}_b \hbox{\rm is symmetric}}
and
\eqn\ {
[ D_i({\overline{G}}^{i.jb}\delta A^\perp _{jb})]^a=0
}
The gauge-fixing conditions that we have discussed so far are good
only locally in superspace and there can be subtleties associated
with Gribov copies[12]. Moreover, the gauge-fixing conditions are
derived for regions in superspace where the supermetric is assumed
to be regular. As we have discussed, this means that we stay away
from singular points with vanishing $\tB$. In the context of perturbation
theory, this restriction may not be unreasonable. Typically, as
in the case of abelian anti-instantons, such configurations have
more symmetry than neighbouring configurations and are thus singular
in the gauge and diffeomorphism-invariant moduli space. The full
quantum theory must of course take into account these intriguing
configurations.

Precisely because the contravariant superspace metric is chosen
to be ${\ut{G}}_{iajb}\equiv {\sep}{}_{ijk}{\sep}{}_{abc}{\widetilde B}^{kc},$
the superhamiltonian constraint can be interpreted as the
{\it free} Klein-Gordon equation in curved superspace with
covariant metric
\eqn\ {
{\widetilde G}^{iajb}\equiv {\widetilde B}^{-1}
\left\{  {{1}\over{2}}{\widetilde B}^{ia}{\widetilde B}^{jb}-{\widetilde
B}^{ib}{\widetilde B}^{ja}
 \right\}
}
This provides a natural ordering for the ``Ashtekar-Wheeler-DeWitt
Equation''
\eqn\ {
 {\widetilde B}^{3/2}   {\widetilde \sigma }^{ia}\left\{{\widetilde B}^{-3/2}
 {\sep_{ijk}}   \sep_{abc}{\widetilde B}^{kc}
\right\}{\widetilde \sigma }^{jb}\Phi [A]   =0
}
which in the connection-representation is equivalent to
\eqn\ {
(\det{\tilde G})^{-{{1}\over{2}}}   {{\delta }\over{\delta A_{jb}}}
(\det{\tilde G})^{{{1}
\over{2}}}{\ut{G}}_{iajb}   {{\delta }\over{\delta A_{ia}}}
\Phi [A]   =0
 }
(The equations are for wavefunctionals which are harmonic zero-forms
in superspace. They can be generalized for instance, to the case
of wavefunctionals of weight ${{1}\over{2}}$ in superspace, by appropriate
insertions of powers of det${\tilde G}$.) Notice that in the classical
context, (39) reduces to
\eqn\ {
H \equiv    \sep_{ijk}   {\sep}{}_{abc}
{\widetilde B}^{kc}{\widetilde \sigma }^{ia}{\widetilde \sigma }^{jb}
 =  0}
and in local coordinates, the Ashtekar-Wheeler-DeWitt Equation can
be written as
\eqn\ {
(  {{\delta }\over{\delta {x^a}_b}}+{\Omega ^b}_a){{\delta }\over{
\delta {x^b}_a}}\Phi [A]   =0
}
where
\eqn\ {
 {{\delta }\over{\delta {x^b}_c}}\Phi    \equiv ({e^c}_b)_{ia}{{
\delta }\over{\delta A_{ia}}}\Phi
}
and
\eqn\ {
 {\Omega ^b}_c\equiv {\widetilde B}^{{{3}\over{2}}}   {{\delta }
\over{\delta A_{ia}}}  \left\{  {\widetilde B}^{-{{3}\over{2}}}({e^b}_c)_{ia}
\right\}
}

The reduced configuration space can be interpreted to be the
{\it light-cone} in curved superspace subject to the gauge-fixing
conditions for SO(3) gauge transformations and 3D diffeomorphisms.
A natural way to order the remaining constraints, which are first
order in the momenta,
is to place the momentum operator on the extreme right
of the constraints, which then read
\eqna\ffive
$$\eqalignno{
( \partial _i{{\delta }\over{\delta A_{ia}}}   +\epsilon _{abc}A_i^b{{
\delta }\over{\delta A_{ic}}}  ) \Phi [A]   &=0  &\ffive a \cr
{}
  \sep_{ijk}{\widetilde B}^{ja}{{\delta
}\over{\delta A^a_k}}   \Phi [A]   &=  0  &\ffive b \cr
}$$
in the $A$-representation. The chosen ordering implies that the wavefunctionals
which satisfy the constraints are invariant under infinitesimal
SO(3) gauge transformations and 3D diffeomorphisms since
\eqn\ {
 \Phi [A+\delta A^g]   =\Phi [A]   +\int    d^3\vec x   \delta A^g_{ia}(\vec x)
 {{\delta }\over{\delta A_{ia}(\vec x)}}\Phi [A]   =\Phi [A]
 }
is ensured by the ordering[13]. Issues related to possible
anomalies in the associated quantum constraint algebra will be
taken up in a future work. It has been argued by others that without
proper regularizations, the closure of the quantum constraint algebra
cannot be addressed meaningfully[14].

With the gauge-fixing conditions and supermetric in hand, we can
now study various limits of the theory and consider perturbations
about backgrounds ($A^*_{ia}$ , ${\widetilde \sigma }_*^{ia}$ )
which satisfy all the constraints. The fields can be decomposed
as
\eqn\ {
 A_{ia}=A^*_{ia}   +   a_{ia}  ,  {\widetilde \sigma }^{ia}=
 {\widetilde \sigma }_*^{ia}+ {\widetilde e}^{ia}
 }
where $a_{ia}$ and ${\widetilde e}^{ia}$ are the fluctuations relative
to the background ($A^*_{ia}$ , ${\widetilde \sigma }_*^{ia}$ ).
To lowest order in the fluctuations, the linearized constraints
are
\eqna\fnine
 $$\eqalignno{
 {D^*}_i{\widetilde e}^{ia}+   {\epsilon ^a}_{bc}a^b_i
{\widetilde \sigma }_*^{ic}
 &=0 &\fnine a \cr
{}
 \sep_{ijk}({\widetilde B}^*)^k_a   {\widetilde e}^{ja}+
 {\widetilde \sigma }_*^{ja}[{D^*}_ia_{ja}-{D^*}_ja_{ia}]   & =0 %
             &\fnine b \cr
}$$
$D^*$ denotes the covariant derivative with respect to the background
gauge connection. Expanding about a background which is compatible
with the constraints, the superhamiltonian constraint can be written
as
\eqna\ffone
$$\eqalignno{
& 2 \sep_{ijk}\epsilon _{abc}{\widetilde B}_*^{kc}
 {\widetilde \sigma }_*^{ia}   {\widetilde e}^{jb}+   \sep_{ijk}
{\widetilde \epsilon }^{klm}
\epsilon _{abc}({D^*}_l a_m)^c{\widetilde \sigma }_*^{ia}
{\widetilde \sigma }_*^{jb}+  &{} \cr
{}
& \sep_{ijk}\epsilon _{abc}{\widetilde B}_*^{kc}
 {\widetilde e}^{ia}   {\widetilde e}^{jb}+2 \sep_{ijk}
{\widetilde \epsilon }^{klm}\epsilon _{abc}({D^*}_l a_m)^c
 {\widetilde \sigma }_*^{ia}   {\widetilde e}^{jb}+  &{} \cr
{}
& \epsilon _{abc}\epsilon ^{cde}a_{id}a_{je}{\widetilde \sigma }_*^{ia}
 {\widetilde \sigma }_*^{jb}  + \hbox{higher order terms} &{} \cr
{}
 &=0  &\ffone{} \cr
}$$

The usual perturbation analysis is to consider fluctuations about
the flat Euclidean $\vec x$-independent background
($A^*_{ia}$ , ${\widetilde \sigma }_*^{ia})$ $=(0$ , $\delta ^{ia}).$
Conventional perturbation analysis is carried out in the metric
or vierbein representation. In the $\ts$ -representation,
the covariant supermetric is the coefficient of the quadratic term
in $A_{ia}$ (the variable conjugate to ${\widetilde \sigma }^{ia}),$
in the superhamiltonian constraint. This covariant supermetric has
the form (det$\widetilde \sigma )^{-{{1}\over{2}}}
({\widetilde \sigma }^{ia}{\widetilde \sigma }^{jb}-$
${\widetilde \sigma }^{ib}{\widetilde \sigma }^{ja})$ which implies
that the metric of superspace parametrized by ${\widetilde \sigma }^{ia}$
takes the form
\eqna\fftwo
$$\eqalignno{
 \delta S^2(\vec x)&=(\det\widetilde \sigma )^{{{1}\over{2}}}({{1}
\over{2}}{\ut{E}}_{ia}{\ut{E}}_{jb}-   {%
\ut{E}}_{ib}{\ut{E}}_{ja})   \delta {\widetilde \sigma }^{ia}
\delta {\widetilde \sigma }^{jb}  &\fftwo a \cr
{}
\quad  &=(\det\widetilde \sigma )^{{{1}\over{2}}}\left\{{{1}\over{6}}({%
\ut{E}}_{ia}\delta {\widetilde \sigma }^{ia})^2- \overline{( {%
\ut{E}}_{ib}\delta {\widetilde \sigma }^{ia}) } \overline{( {%
\ut{E}}_{ja}\delta {\widetilde \sigma }^{jb}) } \right\}%
   &\fftwo b \cr
}$$
where ${\ut{E}}_{ia}$ is the inverse of ${\widetilde \sigma }^{ia}$
and $\overline{({\ut{E}}_{ib}\delta {\widetilde \sigma }^{ia})}$
is the traceless part of (${\ut{E}}_{ib}\delta {\widetilde \sigma }^{ia}).$
Demanding, as before, that the physical modes of the subspace with
supermetric (det$\widetilde \sigma )($ ${{1}\over{3}}{\ut{E}}_{ia}
{\ut{E}}_{jb}-$ ${\ut{E}}_{ib}{%
\ut{E}}_{ja})$ be orthogonal to the gauge directions,
the gauge-fixing conditions are
\eqn\ {
 \epsilon _{abc}{\bar{\tilde e}{   }}^{bc}   =0
 }
and
\eqn\ {
 \partial _{ i}{\bar{\tilde e}{   }}^{ia}   =0
 }
where  ${\bar{\tilde e}{   }}^{ia}$  is the traceless part of $( \delta
{\ts}^\perp )^{ia}$.  The linearized constraints
from (49a) and (49b) are
\eqna\ffive
$$\eqalignno{
 \partial _i{\widetilde e}^{ia}+{{\epsilon ^a}_b}^ia^b_i   &=0 %
     &\ffive a \cr
{}
 \partial _i{a^b}_b-\partial _b{a^b}_i   &=0  &\ffive b \cr
}$$
and to lowest order, the superhamiltonian constraint is
\eqn\ {
 {\epsilon ^{ab}}_c\partial _a{a^c}_b=0
}
The constraints and gauge-fixing conditions are solved by $a_{ia}$
and ${\widetilde e}^{ia}$ being transverse, symmetric and traceless.
These two local degrees of freedom of the theory linearized about
the flat background can be identified with the usual gravitons[16].
In the asymptotically flat context, the boundary Hamiltonian generates
asymptotic time translation and dynamical evolution for the theory.
However, in the case of spatially compact manifolds, the supermetric
(51) suggests that the intrinsic time parameter is proportional
to ln(det$\widetilde \sigma )$ since ${\ut{E}}_{ia}\delta
{\widetilde \sigma }^{ia}=\delta \ln(\det\widetilde \sigma ).$ This
is in agreement with the previous analyses based on the supermetric
(4). In the linearized limit, the intrinsic time coordinate is proportional
to tr($\ts$) since
\eqn\ {
( \delta S)^2\vert _{{\tilde{\sigma}}_*}=   {{1}\over{6}}(
tr \ts )^2-{\bar{\ts}}^{ab}   {\bar{\ts}}_{ba}
}
Keeping the fluctuations to second order in the superhamiltonian
constraint, we have
\eqn\ {
 {a^b}_b{a^c}_c-{a^b}_c{a^c}_b+{\epsilon ^{ab}}_c\partial _a{a^c}_b+4{
\epsilon ^{cb}}_a(\partial _ca_{db}){\widetilde e}^{da}=0
}
As can be expected from the form of superhamiltonian constraint
of the full theory, in the ${\widetilde \sigma}$-representation,
we do not end up with a free Klein-Gordon equation (see also Ref.
15 for connection and loop-representations perturbation analyses
with the flat background). Note that expression (11), the supermetric
of connection-superspace, is singular for this flat Euclidean background
configuration! In the connection-representation, if we wish to consider
the superhamiltonian constraint as the Klein-Gordon Equation
in superspace, perturbing about the flat background is highly
unnatural if not impossible.

We now consider perturbations about an unconventional background
field which will exhibit many of the salient features associated
with the full theory in the connection-representation. Consider
the  $\vec x$-independent  background
\eqn\ {
( A^*_{ia}  ,  {\widetilde \sigma }_*^{ia}) = (\delta_{ia},  0 )
}
This choice of background leads to ${\widetilde B}_*^{ia}=\delta
^{ia}$ and the supermetric takes on a simple form for this configuration.
It is as natural to consider such a background in the $A$-representation
as it is to use the flat background with vanishing Ashtekar connection
in the metric representation with supermetric (3) or in the
$\ts$-representation
with supermetric (51). The background with vanishing ${\ts}^{ia}$
is also an {\it extremum} of superhamiltonian $H$. What is
remarkable about this background is that it is considered to be
unphysical in the context of ADM variables because for vanishing
densitized triads, the ADM variable $g_{ij}$ (considered as derived
from ${\ts}^{ia}$ through $g^{ij}=$ (det$\ts)^{-1}
({\ts}^{ia}{\ts}^j_a))$
is not even well-defined. The supermetric (2) is singular at such
a configuration and the constraints for the ADM variables are not
defined for degenerate metrics. The situation is however very different
with the Ashtekar variables since {\it nowhere} in the
Ashtekar constraints is there a requirement that the conjugate variable
${\ts}^{ia}$ be non-degenerate. Thus it makes perfect
sense to consider the perturbation about zero momenta. Notice also
that this background satisfies all the constraints. Indeed in the
$A$-representation, the condition ${\ts}^{ia}=0$
translates into
\eqn\ {
 {{\delta }\over{\delta A_{ia}}}\Phi [A]   =0
}
which has the interpretation that $\Phi [A]$ is a topological invariant
of $A_{ia}.$ With the ordering of the constraints discussed previously,
the condition (58) is sufficient for all the quantum constraints
to hold and a state which satisfies it is a possible quantum state
of the theory[13]. In particular, a quantum state with this property
is the non-abelian Ray-Singer torsion $\Phi =\int DA$ exp(ik${\cal
C})$ discussed in Ref. 16.

With this unconventional background, the linearized Gauss' Law and
supermomentum constraints take the form of
\eqna\stwo
$$\eqalignno{
 \partial _i{\widetilde e}^{ia} &=  0  &\stwo a \cr
{}
 \sep_{ija}   {\widetilde e}^{ja}   &= 0  &\stwo b \cr
 }$$
Using
\eqn\ {
\left\{  a_{ia}(\vec x)  ,  {\widetilde e}^{jb}(\vec y )
 \right\}_{P.B.}={{1}\over{2}}   {
\delta _a}^b   {\delta _i}^{j}   \delta ^3
( \vec x   -\vec y )
}
the linearized constraints generate
\eqna\sfive
$$\eqalignno{
 \delta {\overline{a}}_{ia}&=\left\{   {\bar a}_{ia},   \int
d^3\vec x  ( \eta _b\partial _j{\widetilde e}^{jb}+\xi ^j \sep_{jkb}
 {\widetilde e}^{kb})\right\}_{P.B.}   &{} \cr
{}
\quad  &={{1}\over{2}}  ( -\partial _i\eta _a+{{1}\over{3}}\delta _{ia}
\partial _b\eta ^b   +\xi ^j\epsilon _{jia}  ) &\sfive{} \cr
}$$
where ${\bar a}_{ia}$ is the traceless part of $a_{ia}.$ Thus the
constraints preserve the tracelessness of ${\bar a}_{ia}$ and suggest
that the longitudinal and anti-symmetric parts of ${\bar a}_{ia}$
can be gauged away. Indeed, with $a_{ia}\equiv \delta A^\perp _{ia},$
the gauge-fixing conditions, (25) and (37), yield
\eqna\ssix
$$\eqalignno{
  {\widetilde \epsilon }^{ija}{\overline{a}}_{ja}&=0  &\ssix a \cr
{}
 \partial _b{{\overline{a}}^b}_a& =0  &\ssix b \cr
}$$
So the constraints and gauge-fixing conditions indicate that the
physical degrees of freedom are in the transverse and symmetric
parts of ${\overline{a}}_{ia}$ and ${\widetilde e}^{ia}.$ With this
background, to lowest order (which is second order since there are
no first order terms because the background with vanishing ${\widetilde \sigma
}^{ia}$
is an extremum of $H$) the superhamiltonian constraint reads
\eqn\ {
  {\widetilde e}^{aa}   {\widetilde e}^{bb}-{\widetilde e}^{ab}
 {\widetilde e}^{ba}=0
}
The perturbations can result in configurations with non-degenerate
metrics when det(${\widetilde e}^{ia})$  is non-vanishing. The two
{\it local} degrees of freedom associated with $a_{ia}$
and ${\widetilde e}^{ia}$ described by Equations (60), (63)-(64),
are the ``nonconventional gravitons''.

All the constraints including the superhamiltonian constraint commute
among themselves. The supermetric at this background configuration
is
\eqn\ {
 {\widetilde G}^{iajb}\vert _{A^*_{ia}}   =  \left\{  {{1}
\over{2}}   \delta ^{ia}\delta ^{jb}   -   \delta ^{ib}\delta ^{ja}
\right\}
}
i.e.
\eqn\ {
( \delta S)^2\vert _{A^*_{ia}}   =  \left\{  {{1}\over{6}}({a^c}_c)^2
 -{{\bar a}^b}_c{{\bar a}^c}_b\right\}
}
where $a_{ia}\equiv \delta A^\perp _{ia}$ and ${\bar a}_{ia}$ is
symmetric, traceless and transverse. In the $a$-representation, as
in the full theory, the superhamiltonian constraint translates into
the {\it free} Klein-Gordon equation
\eqn\ {
\left\{ 6{{\delta ^2}\over{(\delta {a^c}_c)^2}} -{\delta \over
{\delta a_{\alpha}}}{\delta \over{\delta a^{\alpha}}}
\right\}\Phi[{a}]    =0
}
where $ {\alpha} = 1,2,3,+,- $, and the five conjugate variables to
the traceless symmetric components of ${\widetilde e}^{ab}$ can be
written explicitly as
\eqna\deffi
$$\eqalignno{
& a_{1}= \sqrt{2}{a_{(23)}},\qquad a_{2}= \sqrt{2}{a_{(31)}},
\qquad a_{3}= \sqrt{2}{a_{(12)}}, & \deffi a \cr
{}
& a_{+}=\sqrt{3 \over{2}}({\bar a}_{11}+{\bar a}_{22}), \qquad
 a_{-}=\sqrt{1 \over{2}}({\bar a}_{11} - {\bar a}_{22}) & \deffi b \cr
}$$
This suggests that the coordinate $\tau \equiv \sqrt{{{1}\over{6}}}\tr(a)$
should be identified as the intrinsic time in the quantum theory.
The conserved
density can be chosen as the usual one for Klein-Gordon wavefunctionals.
With appropriate restriction to positive frequency modes, it will
be positive definite. ``Plane wave" solutions are given by
exp$({i}\int
{\widetilde e}^{\alpha}a_{\alpha} - tr({\widetilde e})\tau d^3x)$.
However, one can also proceed further and obtain
the massless Dirac equation
\eqn\ {
\left\{ \gamma ^0{{\delta }\over{\delta \tau }} + {\gamma^{\alpha}}
{{\delta }\over{\delta a^{\alpha}}}\right\}\Psi [\tau , a_{\alpha}]
 =0
}
with the conserved ( in $\tau -evolution)$ positive-definite probability
density
\eqn\ {
  \rho =\Psi ^\dagger [\tau ,  \bar{a}]  \Psi [\tau ,  \bar{a}]
 }
It is even possible to contemplate {\it quantum states which are chiral}
in connection superspace.

In momentum space, the two degrees of freedom can be isolated even
more explicitly and the Fourier transform of ${\bar a}_{bc}($ $\vec x)$
is
\eqn\ {
  {\bar a}_{bc}({\vec k})=A^+({\vec k})m_b m_c+A^-({\vec k}){\bar m}_b
{\bar m}_c
}
where (see, for instance, Ref. 16) the basis vectors satisfy
\eqn\ {
 m_am^a={\bar m}_a{\bar m}^a=k_am^a=k_a{\bar m}^a=0,\qquad
m_a{\bar m}^a=1
}
and ${\bar m}_a$ is the complex conjugate of $m_a.$ The physical
modes $A^\pm $ are the positive and negative helicity modes.
The same decomposition can be done for the transverse, symmetric
and traceless conjugate variable ${\widetilde e}^{bc}$.
As stated in the beginning, we have concentrated on real Ashtekar
variables and this is consistent for spacetimes with Euclidean
signature. The reality of $a_{ab}($ $\vec x)$ (which need to be
imposed only on the physical modes) is equivalent to the condition
\eqn\ {
[\overline{a_{bc}({\vec k})}] =a_{bc}(-{\vec k})
}
Similar reality conditions can be imposed on the physical modes
of the conjugate variables.

It has been postulated by many authors that in quantum gravity,
the signature of spacetime is not sacred and fluctuations of it
can occur. Certainly for fluctuations about the background at which
the metric is not even defined, it is rather unnatural to the impose
a set of reality conditions on the Ashtekar variables to restrict the
configurations to correspond to spacetimes with Lorenztian signature.
It is more natural to start with complex variables and demand the
wavefunctionals to be holomorphic in the Ashtekar potentials[6,7].
Although it may no longer be true that for complex potentials,
the signature of the supermetric is as in expression (17), the
decomposition (19) can still be carried out and the gauge-fixing
conditions will not be altered. If we start with the Ashtekar variable
written as $A_{ia}=iK_{ia}-{{1}\over{2}}\epsilon _{abc}\omega _i^{bc},$
it can be checked that provided the other constraints are satisfied,
despite ${\widetilde B}^{ia}$
being complex, the superhamiltonian constraint (5c) remains real
if ${\widetilde \sigma }^{ia}$ is real. This suggests that in the quantum
theory, despite the complex Ashtekar potentials, we should require
the Laplacian operator in superspace to be hermitian with respect to a
suitable measure and inner product after gauge-fixing.

We emphasize that the ``unconventional'' background is inaccessible
to conventional perturbation analyses with the ADM variables and
cannot be perturbatively related to the flat background. This highly
interesting limit of the theory is precisely the zero-momentum limit
of quantum gravity with Ashtekar variables. Various interesting
questions such as the perturbative renormalizabilty (or non-renormalizability)
of the theory about this unconventional background, the physical
implications of spin and chiral quantum states of gravity, the influence
of matter fields on the stability of the theory, and the intriguing
role of configurations with vanishing $\tB$ in the dynamics of the
full theory immediately come to mind, and are being studied. We
hope to address these issues in a future report.

\bigskip
\bigskip
\bigskip
The research for this work is supported in part by the DOE under Grant No.\
DE-FG05-92ER40709. One of us (CS) acknowledges the support of a Cunningham
Fellowship.
\vfil\eject
\bigskip
\centerline{\bf References}
\bigskip

\item{[1]}R. Arnowitt, S. Deser and C. Misner, in {\it Gravitation:
An introduction to Current Research}, ed. L. Witten (Wiley,
New York, 1962); Phys. Rev. {\bf 117}, 1595 (1960).
\item{[2]}M. Henneaux, M. Pilati and C. Teitelboim, Phys. Lett. {\bf 110B
}, 123 (1982); M. Pilati, Phys. Rev. {\bf D26}, 2645
(1982); ibid. {\bf D28}, 729 (1983).
\item{[3]}B. S. DeWitt, Phys. Rev. {\bf 160}, 1113 (1967).
\item{[4]}A. Ashtekar, Phys. Rev. Lett. {\bf 57}, 2244 (1986).
\item{[5]}A. Ashtekar, Phys. Rev. {\bf D36}, 1587 (1987).
\item{[6]}A. Ashtekar, in {\it New Perspectives in Canonical Gravity
}(Bibliopolis, Naples, 1988).
\item{[7]}A. Ashtekar, {\it Lectures on Non-perturbative Canonical Gravity
}, Advanced Series in Astrophysics and Cosmology-Vol. 6. (
World Scientific, Singapore, 1991).
\item{[8]}K. Kuchar, {\it Canonical Quantum Gravity}, gr-qc/9304012.
\item{[9]}L. N. Chang and C. P. Soo, {\it Einstein Manifolds in Ashtekar
Variables: Explicit Examples}, hep-th/9207056, VPI-IHEP-92-5;
C. P. Soo, {\it Classical and Quantum Gravity with Ashtekar Variables
}. Ph. D. Thesis, VPI\&SU (1992), VPI-IHEP-92-11.
\item{[10]}G. B. Gibbons and S. W. Hawking, Phys. Rev. {\bf D15},
2752 (1977).
\item{[11]}P. Mazur and E. Mottola, Nucl. Phys. {\bf B341}, 187
(1990); Z. Bern, S. K. Blau and E. Mottola, Phys. Rev. {\bf D43
}, 1212 (1991); P. Mazur, Phys. Lett. {\bf 262B},
405 (1991).
\item{[12]}V. N. Gribov, Nucl. Phys. {\bf B139}, 1 (1978); I.
M. Singer, Comm. Math. Phys. {\bf 60}, 7 (1978).
\item{[13]}T. Jacobson and L. Smolin, Nucl. Phys. {\bf B299},
295 (1988).
\item{[14]}J. L. Friedman and I. Jack, Phys. Rev. {\bf D37}, 3495
(1987); N. C. Tsamis and R. P. Woodard, Phys. Rev. {\bf D36},
3641 (1987).
\item{[15]}A. Ashtekar, C. Rovelli and L. Smolin, Phys. Rev.
{\bf D44}, 1740 (1991).
\item{[16]}E. Witten, Comm. Math. Phys. {\bf 121}, 351 (1989).

\bye